\documentclass[manuscript]{aastex631_arxiv}

\usepackage{color}
\usepackage{graphicx}
\usepackage{bm}
\usepackage{epsf}
\usepackage{wrapfig}
\usepackage{multirow}
\usepackage{epstopdf}
\usepackage{tabularx}
\usepackage{amssymb,amsmath}

\shorttitle{solar cycle}
\shortauthors{Huang et al.}

\begin{document}

\title{Adjusting the Potential Field Source Surface Height Based on MHD Simulations}

\correspondingauthor{Zhenguang Huang}
\email{zghuang@umich.edu}

\author[0000-0003-1674-0647]{Zhenguang Huang}
\affiliation{Climate and Space Sciences and Engineering, University of Michigan, Ann Arbor, MI 48109, USA}

\author[0000-0001-8459-2100]{G\'abor T\'oth}
\affiliation{Climate and Space Sciences and Engineering, University of Michigan, Ann Arbor, MI 48109, USA}

\author[0000-0002-9954-4707]{Jia Huang}
\affiliation{Space Sciences Laboratory, University of California, Berkeley, CA 94720, USA.}

\author[0000-0001-9114-6133]{Nishtha Sachdeva}
\affiliation{Climate and Space Sciences and Engineering, University of Michigan, Ann Arbor, MI 48109, USA}

\author[0000-0001-5260-3944]{Bart van der Holst}
\affiliation{Climate and Space Sciences and Engineering, University of Michigan, Ann Arbor, MI 48109, USA}

\author[0000-0003-0472-9408]{Ward B. Manchester}
\affiliation{Climate and Space Sciences and Engineering, University of Michigan, Ann Arbor, MI 48109, USA}

\begin{abstract}

A potential field solution is widely used to extrapolate 
the coronal magnetic field above the Sun's surface to a certain
height. This model applies the current-free approximation 
and assumes that the magnetic field is entirely radial beyond the source
surface height, which is defined as the radial distance from the center of the Sun. 
Even though the source surface
is commonly specified at 2.5 $R_s$ (solar radii), previous
studies have suggested that this value is
not optimal in all cases. In this study, we propose a novel
approach to specify the source surface height, by
comparing the areas of the open magnetic field regions from the
potential field solution with predictions made by a magnetohydrodynamics model, in our case the Alfv\'en Wave Solar atmosphere Model. 
We find that the adjusted source surface height is significantly less
than 2.5 $R_s$ near solar minimum, and slightly 
larger than 2.5 $R_s$ near solar maximum. We also report that the adjusted
source surface height can provide a better open flux agreement
with the observations near the solar minimum, while the 
comparison near the solar maximum is slightly worse.

\end{abstract}

\section{Introduction}

The magnetic field configuration in the solar corona is critical to understanding
the physics of the solar corona as it is highly associated with the origin of the
fast and slow solar wind according to several theories including the flux expansion model \citep{Wang_1990,Arge_2000} and the separatrix–web (S-web) model \citep{Antiochos_2011,Titov_2011}. The coronal magnetic structure defines the structure of the entire heliosphere including the location of the helispheric current sheet \citep{McComas_2002}, the distribution of ion charge state \citep{Oran_2013}, and the formation and evolution of stream interaction regions \citep{Gosling_1999}

There are different approaches in constructing the magnetic field in the solar corona. For example, to obtain the global coronal structures, the potential field source surface model \citep{Altschuler_1969,Schatten_1969} and non-force free models \citep{Bogdan_1986, Neukirch_1995, Hu_2008a} are applied. Besides, more sophisticated Magnetohydrodynamics (MHD) models 
such as the Magnetohydrodynamic Algorithm outside a Sphere 
(MAS, \cite{Mikic_1999}), the hybrid solar wind model of \cite{Feng_2011} and the Alfv\'en Wave Solar atmosphere Model (AWSoM, \cite{Sokolov_2013,vanderholst_2014})
are also used. For active regions, linear force-free models \citep{Alissandrakis_1981, Gary_1989}, non-linear force-free models \citep{Schrijver_2006,Aschwanden_2013} and non-force free models \citep{Hu_2008b} are typically used.

The Potential Field Source Surface 
(PFSS, \cite{Altschuler_1969,Schatten_1969}) model is the simplest model and 
widely used for extrapolating the magnetic field in the solar corona 
based on the observed photospheric magnetic field. It assumes a current-free
magnetic field configuration in the solar corona. Under this assumption, 
the magnetic field equation can be simplified as the Laplace equation as
$\nabla^2 \phi = 0$, where $\phi$ is the scalar potential of the magnetic field. 
The inner boundary is usually specified by the observed photospheric magnetic
field (a magnetogram); and a purely radial magnetic field is assumed 
at the outer boundary, which is also called the source surface and 
usually at 2.5 $R_s$ (solar radii). The magnetic field can be obtained as
$\mathbf{B} = \nabla \phi$ after solving the Laplace equation.

The simple PFSS model has very limited adjusted parameters: the height and or the shape of the source surface. For the traditional spherical shape of the source surface, \cite{Altschuler_1969} have suggested that the value
is within 1.5 to 4.0\,$R_s$. They also compared the magnetic
field structures with white-light coronagraph observations and concluded
that the source surface of 2.5\,$R_s$ best represented the overall structures. \cite{Lee_2011} used
the synoptic photospheric magnetic field map from the Mount Wilson Observatory (MWO)
and investigated the source surface height by comparing the observed 
coronal holes 
and the interplanetary field strength as well as polarity measurements 
during minimum periods in the solar cycles 22 and 23, and concluded that the optimal
height is between 1.5 to 1.9\,$R_s$, which is significant less than
the typically used value of 2.5\,$R_s$. Recently, \cite{Nikolic_2019} 
derived the open solar magnetic flux and coronal holes for various heights of the
PFSS model and found that source surface was significantly lower than 2.5\,$R_s$
during the active phase of solar cycle 24. 
\cite{Badman_2020} 
explored the source surface height and suggested an extraordinarily low source 
surface height (1.3 - 1.5\,$R_s$) can predict the small-scale polarity inversions 
observed by the first Parker Solar Probe (PSP) encounter between 
October 15 and November 30 in 2018. On the other hand, the non-spherical shape
of the source surface was proposed in early 1980s \citep{Levine_1982}. A follow-up study by \cite{Schulz_1997} also investigated the non-spherical
source surface which was formulated with $F = r^{-k}\tilde{B}$, where $r$ is
the heliocentric distance and $\tilde{B}$ is the scalar magnitude of the magnetic
field produced by currents inside the Sun. They found that $k \approx 1.4$ was
a good choice when they compared their results with MHD solutions for a single rotation between May and June 1993. More recently, \cite{Kruse_2020} developed a numerical solver for an elliptic source surface of the PFSS model and investigated the different magnetic field configurations between the spherical and elliptical source surfaces. \cite{Kruse_2021} mapped the in situ solar wind observations from the Advanced Composition Explorer (ACE) and Solar TErrestrial Relations Observatory (STEREO) back to the source surface with the ballistic mapping technique, and compared the predicted magnetic field polarity at the source surface.
They showed that the PFSS model with oblate elliptical source surfaces elongated along the solar equatorial plane performed slightly better than the traditional spherical shape. They also suggested that the non-spherical shape of the source surface was more important in solar minimum than in solar maximum.

The source surface height can also affect the predicted
solar wind speed and density with the Wang-Sheeley-Arge (WSA) model \citep{Wang_1990, Wang_1992, Arge_2000}, one of the widely used semi-empirical solar wind models in the community. 
\cite{Meadors_2020} used data assimilation with
particle filtering (sequential Monte Carlo) to adjust the height of the source
surface and found that solar wind predictions can improve by varying the height of the source surface. On the other hand, 
\cite{Issan_2023} argued that the source surface height 
does not play a critical role in changing the simulated solar
wind with their Bayesian inference and global sensitivity analysis, compared
to other parameters of the WSA model. However, their analysis was limited
to three Carrington Rotations (CRs) during the declining
phase of solar cycle 23, and it is unclear whether their conclusion is valid
during the entire solar cycle.

In this study, we propose a novel approach in adjusting the source
surface height by comparing the magnetic field structures from the PFSS and AWSoM
models. Section \ref{sec:method} briefly describes the PFSS and AWSoM models. Section
\ref{sec:results} discusses how we adjust the source surface height based on AWSoM
simulation results. Finally, 
Section \ref{sec:summary} draws our conclusions. 

\section{Methodology} 
\label{sec:method}

The PFSS model assumes that the magnetic field is current-free
in the solar corona between the solar surface and the source surface. 
\cite{Toth_2011} proposed two approaches to obtain the
potential field solution from the Laplace equation: 
1. a spherical harmonics expansion based on a uniform
latitude grid; 2. a finite difference solver. We use the spherical harmonics expansion
and choose the degree of spherical harmonics to be 180 in this study based on the 1 degree longitudinal resolution of the magnetogram. 
The spherical harmonics solution is evaluated on a uniform spherical grid with 150, 180 and 360 cells in the radial, latitudinal and longitudinal directions, 
respectively. The radial coordinate extends from the solar surface at $r=1\,R_s$. The source surface height is varied between $1.5$ and $3.5$\,$R_s$ to
explore how this parameter impacts the open field area as well as the open flux, 
while the default source surface height is 2.5\,$R_s$.
In order to directly compare with the MHD solutions in \cite{Huang_2023}, we use 
the same ADAPT-GONG magnetograms as inputs, which are listed in 
Table\,\ref{tab:results}. 

For our study, the coronal magnetic fields are provided by the AWSoM simulations, which are described in \cite{Huang_2023}. AWSoM is 
a first-principle based MHD model, which assumes that 
the nonlinear dissipation and pressure gradient from the 
Alfv\'en wave turbulence is the only source
to drive the solar wind and heat the solar corona. It 
is implemented in the BATS-R-US (Block Adaptive Tree Solar Wind Roe-type
Upwind Scheme) code \citep{Groth_2000, Powell_1999} within the Space Weather Modeling Framework (SWMF) \citep{Toth_2005, Toth_2012, Gombosi_2021}. 
There is only one observational data input for the model: the observed radial component of the photospheric 
magnetic field at the inner boundary. 
At the inner boundary, a uniform density (n=$2\times10^{17}$\,m$^{-3}$) 
and temperature (T=50,000\,K) distribution
is specified. And at the outer boundary, a zero gradient condition is
applied so that the solar wind can freely leave the simulation domain.
\cite{Sokolov_2013} and \cite{vanderholst_2014} described the physics within AWSoM
in great detailed while \cite{vanderholst_2022} discussed the recent improvement 
of the Alfv\'en wave turbulence cascade.
AWSoM results have shown reasonable agreement with in-situ and remote observations under
different solar wind conditions \citep{Jin_2012, Oran_2013, vanderholst_2019,
Sachdeva_2019, Sachdeva_2021, Huang_2023, vanderholst_2022,Szente_2022,Szente_2023,Shi_2024}
and widely used in the community \citep{Jian_2016, Lloveras_2020, Henadhira_2022}. 

\section{Adjusting the Source Surface Height} 
\label{sec:results}

\begin{figure}[ht!]
\center
\gridline{\fig{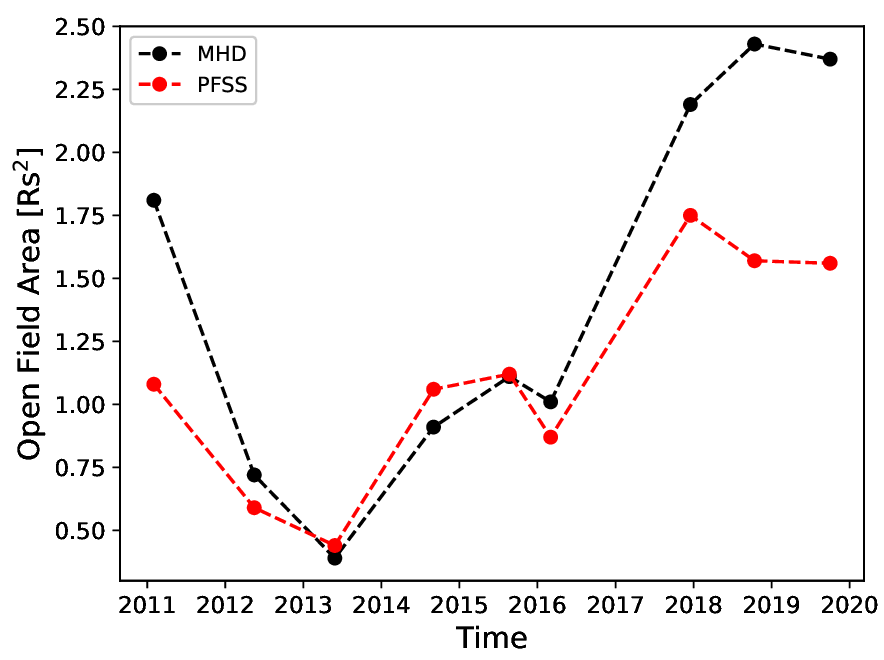}{0.47\textwidth}{(a)}
          \fig{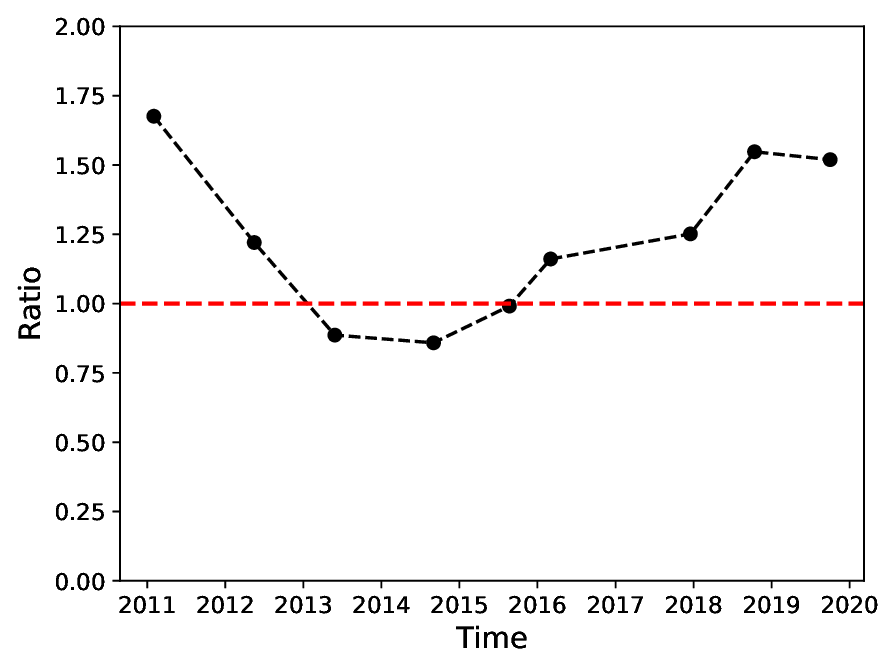}{0.47\textwidth}{(b)}}
          
\caption{Panel (a) shows the open field areas obtained from the optimal AWSoM 
simulations (in black) and the PFSS model (in red) 
for all the Carrington rotations simulated in 
Table\,\ref{tab:results},
while Panel (b) plots the ratio between the AWSoM
and the PFSS open field areas. The horizontal red dashed line indicates
the ratio of 1.}
\label{fig:area}
\end{figure}

We first compare the open field areas obtained from the PFSS and AWSoM models
with the commonly used source surface height as 2.5 R$_s$, 
which are shown in Figure\,\ref{fig:area}. The AWSoM results are directly
obtained from \cite{Huang_2023}. It can be readily noticed that
the open field areas from the AWSoM simulations (in black) 
tend to give larger values
near the solar minimum and slightly 
smaller values near the solar maximum, than
the PFSS results (in red). 
The ratio between the open field area from
the MHD solution and the PFSS solution can be as large as 1.75 for CR2106 (near solar
minimum) and as small as 0.8 for CR2154 (near solar maximum).

\begin{figure}[ht!]
\center
\gridline{\fig{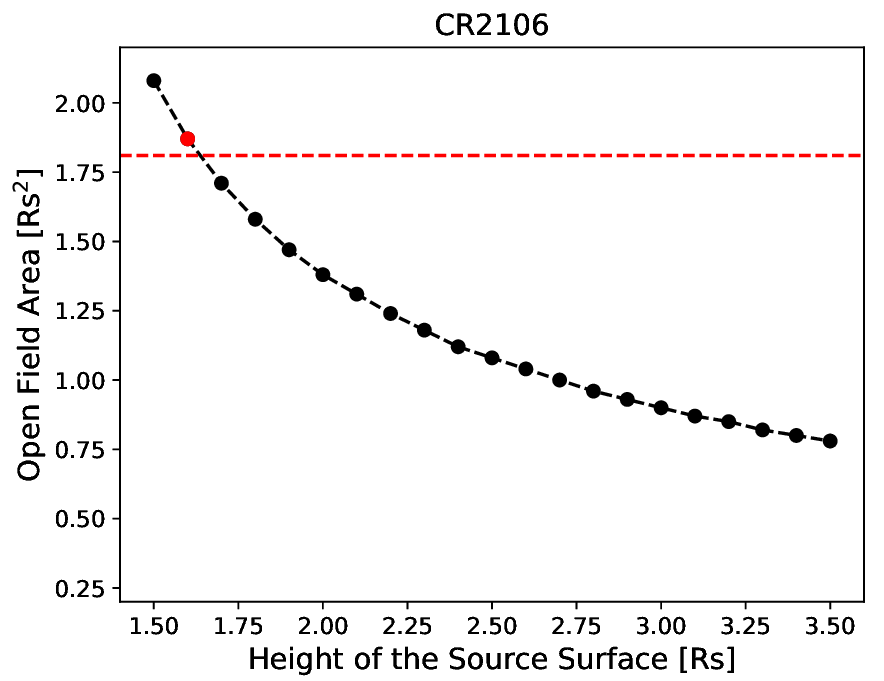}{0.47\textwidth}{(a)}
          \fig{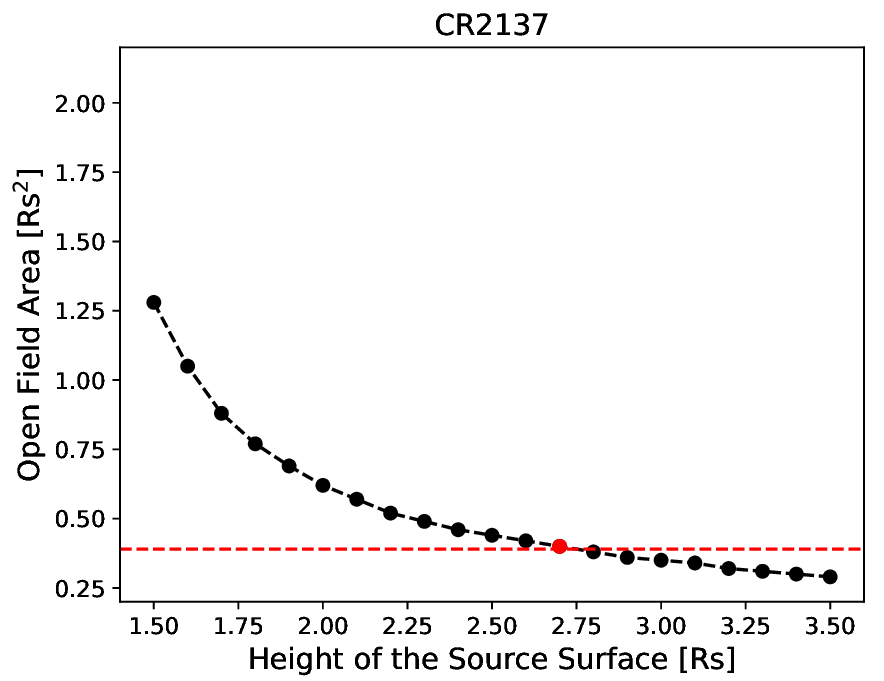}{0.47\textwidth}{(b)}}
          
\caption{The open field area variations with different source surface heights for
CR2106 (Panel (a)) and CR2137 (Panel(B)), respectively. The red spheres show
the open field areas obtained from the adjusted source surface and 
the horizontal red dashed lines indicate the open field areas from the AWSoM
results. 
}
\label{fig:pfss}
\end{figure}

A natural question is, can the source surface height of the PFSS model
be inferred from AWSoM simulations? To be specific, the source surface height 
for the PFSS model is adjusted so that it gives a similar open field area as the AWSoM results.
In order to explore this idea, 
we obtain the PFSS solutions with different heights of the source 
surface, between 1.5 and 3.5 R$_s$ with the step size as 0.1 R$_s$. 
We also derive the open and closed regions by tracing magnetic field lines on 
a refined grid of every 0.5 degree by 0.5 degree in both latitudinal 
and longitudinal directions at 1.01 R$_s$ to avoid tiny closed loops. 
Figure\,\ref{fig:pfss} shows how the open field area
changes with different heights of
the source surface for CR2106 near solar minimum and CR2137 near solar maximum. 
Based on Figure\,\ref{fig:pfss}, we conclude that the open field area monotonically
increases as the source surface height decreases. 
And in order to obtain similar open field areas as
AWSoM, the adjusted source surface heights are 1.6 R$_s$ and 2.7 R$_s$ 
for CR2106 and CR2137, respectively.

\begin{figure}[ht!]
\center
\gridline{\fig{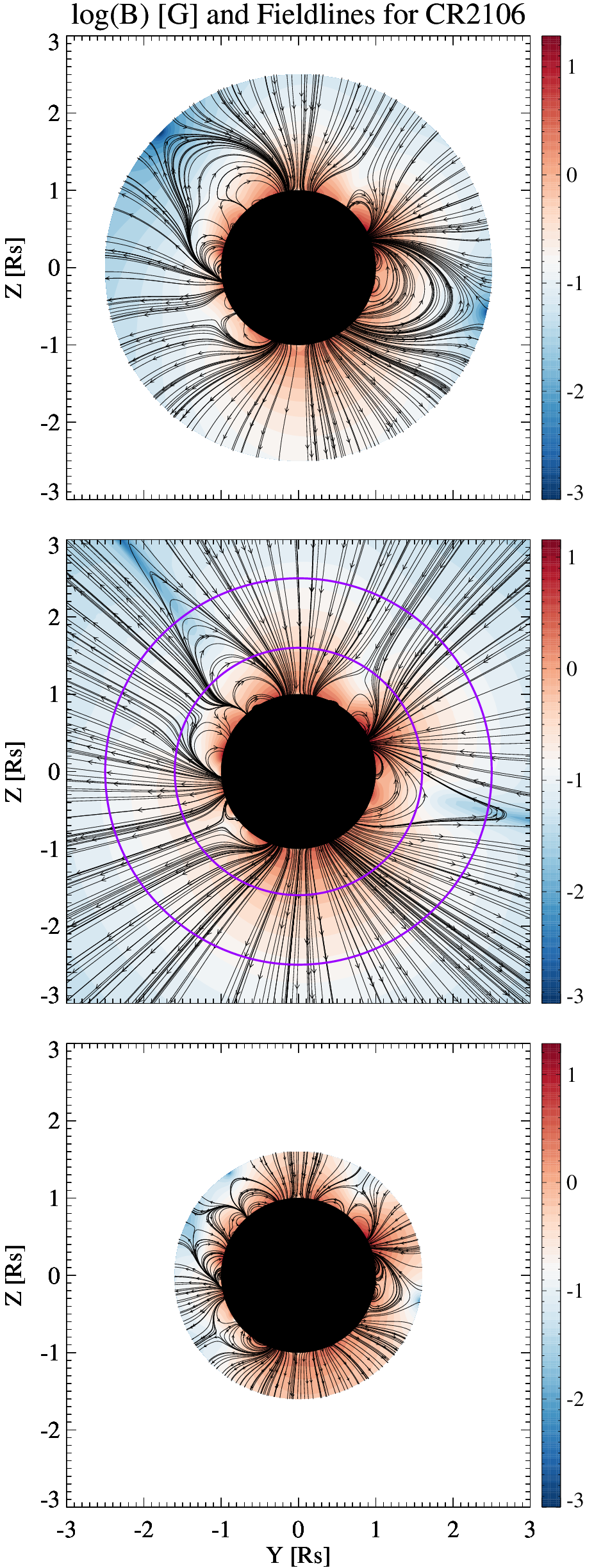}{0.4\textwidth}{(a)}
          \fig{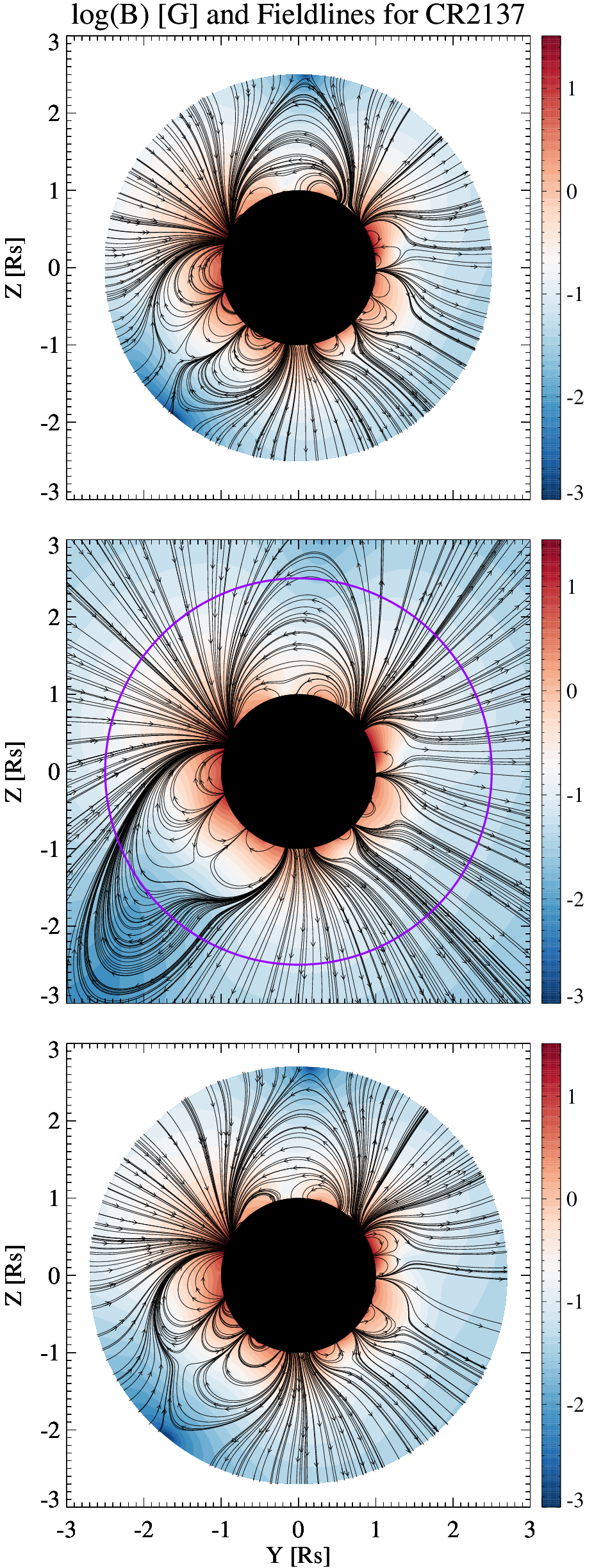}{0.4\textwidth}{(b)}}
          
\caption{The magnetic field configurations from the PFSS model with the default SS height (top
panels) and the adjusted source surface height (bottom panels), as well as from the AWSoM
simulations (middle panels). The color bars show the magnetic field magnitude in log scale. Column (a) shows the results for CR2106 while Column (b) plots the results for 
CR2137, respectively. In Panel (a), the outer purple circle indicates the default source surface height of 2.5\,R$_s$ while the inner purple circle is associated with the adjusted
source surface height of 1.6\,R$_s$. In Panel (b), the purple circle indicates the default source surface height of 2.5\,R$_s$.}
\label{fig:x=0}
\end{figure}

The next step is to examine the magnetic field structures between the PFSS and
AWSoM results. Figure\,\ref{fig:x=0} shows the magnetic field configurations
on the $X=0$ plane for the PFSS model (with the default and adjusted source surface
heights) and the AWSoM results. If we only consider the magnetic field structures
within the PFSS domain with the default source surface height of 2.5\,$R_s$, then the magnetic field topology of the
PFSS model is very different from the AWSoM solution for CR2106, 
especially in the lower right and upper left sides where the streamers
form. The PFSS fieldlines are more rounded at larger height as 
compared to the AWSoM fieldlines. However, if the source surface 
height is adjusted to 1.6 R$_s$,
the magnetic field topology in the PFSS domain
is similar to the AWSoM configuration.
We conclude that the magnetic field topology of the adjusted source surface
height is closer to the AWSoM result if we only consider the topology in the PFSS domain.
For CR2137, as the adjusted source surface height is very close to the default height, 
the differences between the two PFSS solutions are not significant, 
and both of them are in reasonable agreement with the AWSoM result.

\begin{figure}[ht!]
\center
\includegraphics[width=0.9\linewidth]{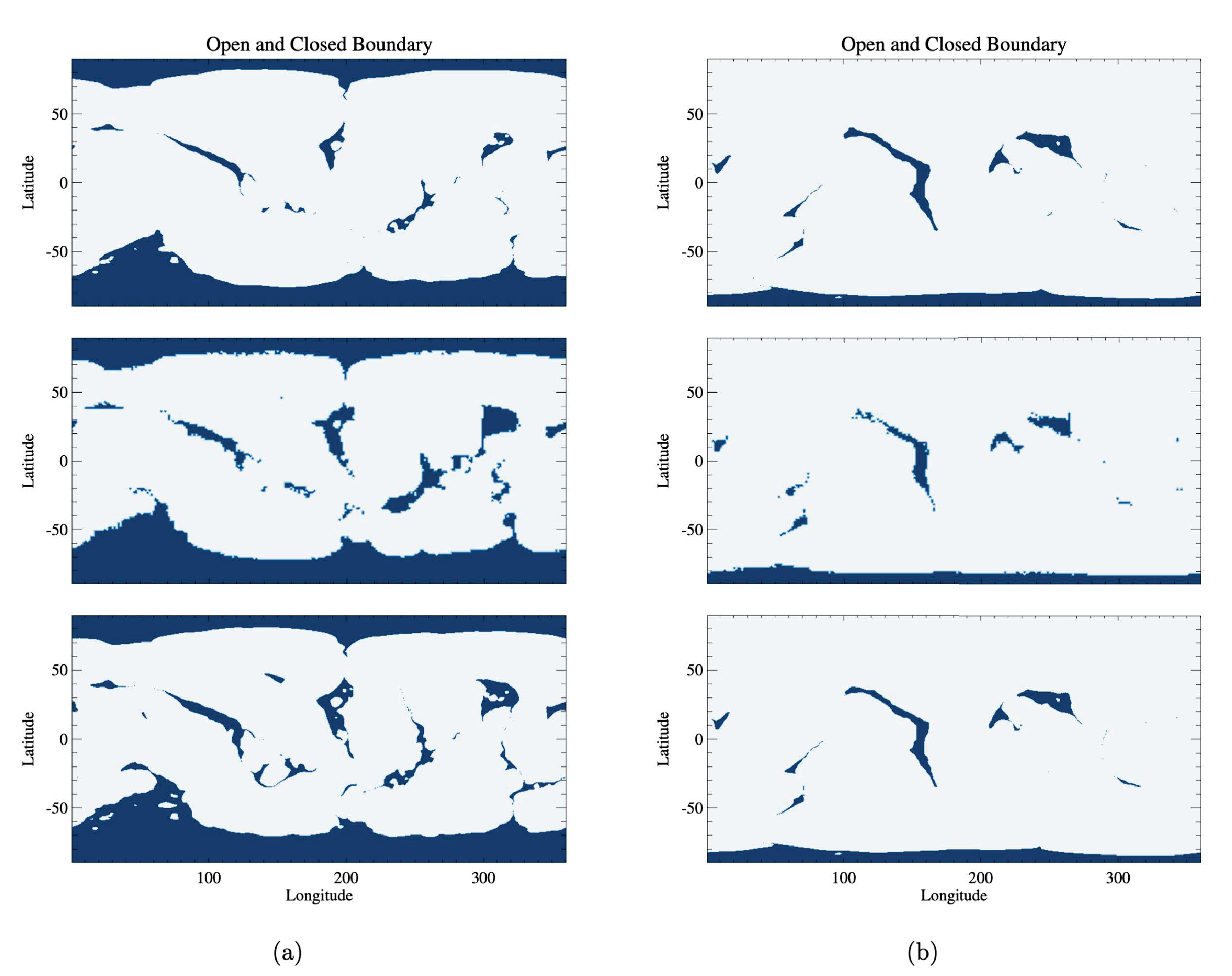}
          
\caption{The blue areas show the open field regions from the PFSS model 
with the default SS height (top
panels) and the adjusted SS height (bottom panels), as well as from the AWSoM
simulation (middle panels). 
Column (a) shows the results for CR2106 while Column (b) plots the results for 
CR2137, respectively.}
\label{fig:boundary}
\end{figure}

Figure\,\ref{fig:boundary} compares the open field regions (at 1.01\,$R_s$) obtained from the PFSS model with
the default and adjusted source surface heights as well as the AWSoM results. For CR2106 near solar minimum, 
we can immediately see the differences between the
two source surface heights, and the open field regions
obtained from the adjusted source surface height is closer to the AWSoM solution. 
For CR2137 near solar maximum, the differences between the two 
source surface heights are not significant and can hardly be 
distinguished visibly; and fortunately, 
both PFSS model results are similar to the AWSoM results.

We carry out the same analysis for other CRs in \cite{Huang_2023} and 
the results are summarized in Table\,\ref{tab:results}. 
The adjusted source surface height is below 2.0 R$_s$ near solar minimum conditions
while it is slightly above 2.5 R$_s$ near solar maximum conditions. In addition, 
we calculate the open flux, which is the 
integral of the unsigned radial component of the magnetic field ($B_r$)
in the open field regions. Table\,\ref{tab:results} shows that 
the open flux increases as the source surface height is reduced. 
Moreover, we calculate the open flux from Wind observations at 1
AU based on the equation $4 \pi R_{1 AU}^2|B_{r,1 AU}|$ 
\cite[e.g.][]{Wang_2000,Linker_2017}, 
where $R_{1 AU} = 215 \ R_s$ and $|B_{r,1 AU}|$ is the 
absolute value of the radial magnetic field strength at 1 AU. 
The open fluxes from the PFSS model with adjusted source surface heights
are closer to AWSoM results and the observations, 
except for CR2137 and CR2154 near solar maximum.

\begin{table}[]
\scriptsize
\begin{tabular}{|c|c|c|cccl|}
\hline
\multirow{2}{*}{CR} & \multirow{2}{*}{UTC Time of Map} & 
\multirow{2}{*}{Adjusted Source Surface Height [$R_s$]} & \multicolumn{4}{c|}{Open Flux [Gauss $R_s^2$]}                                                                          \\ \cline{4-7} 
                                     &                                  &                                       & \multicolumn{1}{c|}{AWSoM}   & \multicolumn{1}{c|}{Default PFSS}  & \multicolumn{1}{c|}{Adjusted PFSS} & Observed \\ \hline
2106                                 & 2011-2-2  02:00:00               & 1.6                                   & \multicolumn{1}{c|}{9.39}  & \multicolumn{1}{c|}{4.83}  & \multicolumn{1}{c|}{9.67}          & 13.76    \\ \hline
2123                                 & 2012-5-16 20:00:00               & 2.2                                   & \multicolumn{1}{c|}{7.99}  & \multicolumn{1}{c|}{5.15}  & \multicolumn{1}{c|}{6.45}          & 13.84    \\ \hline
2137                                 & 2013-5-28 20:00:00               & 2.7                                   & \multicolumn{1}{c|}{5.88}  & \multicolumn{1}{c|}{5.88}  & \multicolumn{1}{c|}{5.32}          & 13.74    \\ \hline
2154                                 & 2014-9-2 20:00:00                & 2.7                                   & \multicolumn{1}{c|}{10.07} & \multicolumn{1}{c|}{9.61}  & \multicolumn{1}{c|}{8.47}          & 16.17    \\ \hline
2167                                 & 2015-8-23 02:00:00               & 2.5                                   & \multicolumn{1}{c|}{15.18} & \multicolumn{1}{c|}{13.36} & \multicolumn{1}{c|}{13.36}         & 17.37    \\ \hline
2174                                 & 2016-3-3 02:00:00                & 2.3                                   & \multicolumn{1}{c|}{11.13} & \multicolumn{1}{c|}{8.51}  & \multicolumn{1}{c|}{9.6}           & 14.71    \\ \hline
2198                                 & 2017-12-17 02:00:00              & 1.9                                   & \multicolumn{1}{c|}{11.45} & \multicolumn{1}{c|}{8.31}  & \multicolumn{1}{c|}{10.58}         & 15.22    \\ \hline
2209                                 & 2018-10-13 06:00:00              & 1.7                                   & \multicolumn{1}{c|}{12.32} & \multicolumn{1}{c|}{7.88}  & \multicolumn{1}{c|}{10.89}         & 12.12    \\ \hline
2222                                 & 2019-10-2 02:00:00               & 1.7                                   & \multicolumn{1}{c|}{11.85} & \multicolumn{1}{c|}{7.93}  & \multicolumn{1}{c|}{10.55}         & 14.41    \\ \hline
\end{tabular}
\caption{The adjusted source surface heights and open fluxes for all the ADAPT-GONG 
magnetograms.}
\label{tab:results}
\end{table}

\section{Summary and Discussions}
\label{sec:summary}

We propose a novel approach to adjust the source surface height of the PFSS model
by matching the size of the open field area to an MHD model, e.g., AWSoM. 
The PFSS model is widely used in the community to study the
magnetic field structures of the solar corona. The source surface height is regularly
set at 2.5\,$R_s$. On the other hand, there are studies arguing this ``default" value may
not be the optimal value \citep{Lee_2011, Nikolic_2019, Badman_2020}. 
The magnetic field structures obtained from an MHD model
are considered to be more accurate than the PFSS model, 
as the MHD model does not employ specific assumptions 
(i.e., current-free and radial field beyond the source surface) to extrapolate 
the magnetic field from the observed photospheric
magnetic field. 
Our new approach connects the magnetic field topology from
the PFSS and AWSoM results. 
Our results suggest that the source surface height is smaller 
than the default height of 2.5 R$_s$ during solar minimum, while it is 
slightly larger than the 2.5 R$_s$ during solar maximum.
We also find that the PFSS model with the
adjusted source surface height can give similar
open field regions as the AWSoM results provided in \cite{Huang_2023}.  

\begin{figure}[ht!]
\center
\includegraphics[width=0.9\linewidth]{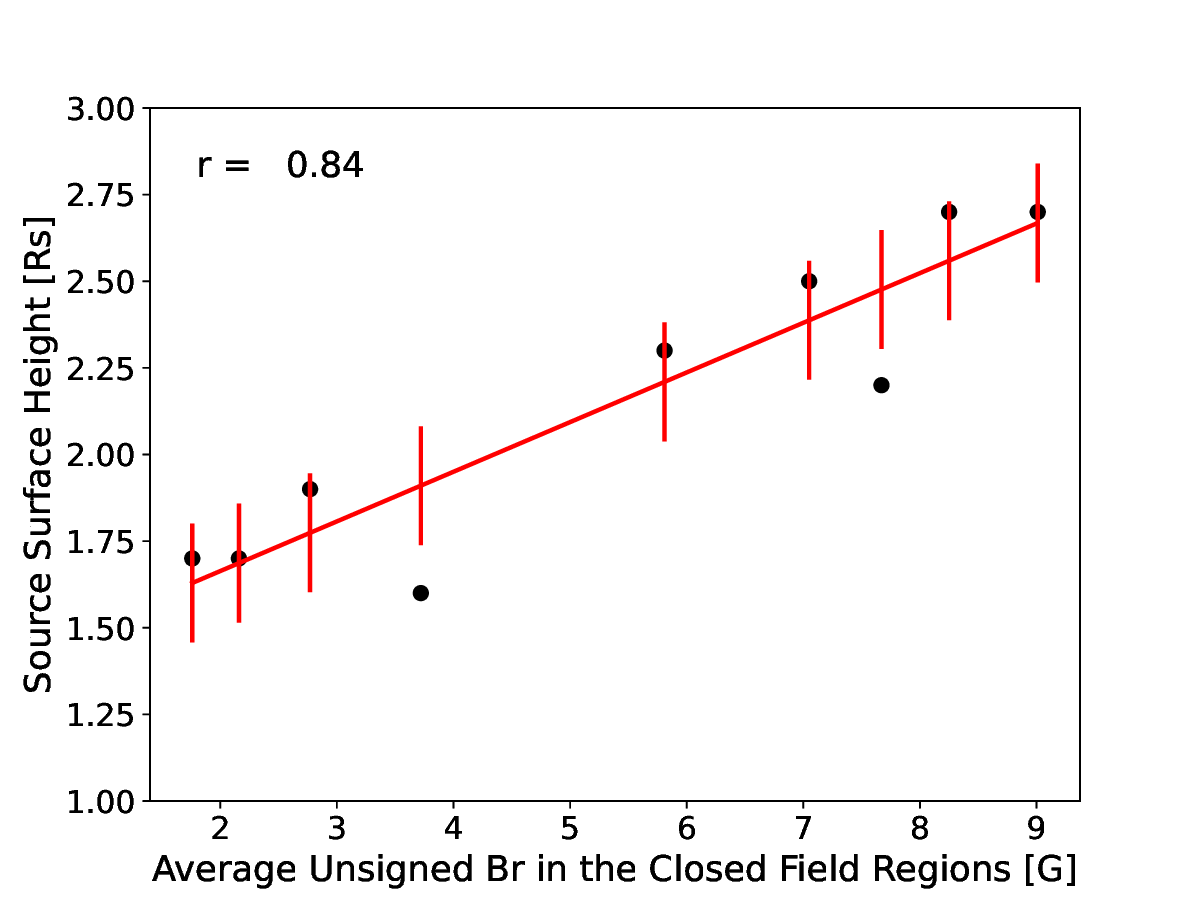}
\caption{The black circles plot the adjusted source surface heights along with the average unsigned $B_r$ in the closed field regions and the red line shows the linear regression between them, where the error bars show the standard deviation of the linear regression. The Spearman’s correlation is shown in the
upper left corner. 
}
\label{fig:reg}
\end{figure}

We compare the magnetic field structures on $X=0$ plane from the 
AWSoM simulations and the PFSS solutions with different source surface heights. 
We notice that the magnetic field structures from the 
adjusted source surface height are closer to the AWSoM configurations, 
if we only consider the topology within the PFSS domain,
especially during solar minimum. In the solar minimum conditions, 
there are very few active regions.
Due to the solar wind acceleration, 
the magnetic fields are dragged in the radial
direction, as seen in the helmet streamers in the middle panels in Figure\,\ref{fig:x=0}, which
may explain the reduced source surface height.
In the solar maximum conditions, 
there are very strong active regions, which bring 
large coronal loops over those regions. 
In this case, the source surface height needs 
to increase to capture the large coronal loops over the active regions. 
We further determine that the Spearman’s correlation coefficient between the 
adjusted source surface height~($r_{adjusted}$) and the average 
unsigned $B_r$ in the closed field regions (at 1.01\,$R_s$ where we determine the open and closed regions from the PFSS solution with the source surface height of 2.5\,$R_s$)
is 0.84. Using the linear regression method, we then find that they can be empirically formulated as:
\begin{equation}
r_{adjusted} = 0.14  \cdot |B_r| + 1.38 \pm 0.17
\end{equation}
where the average unsigned $B_r$ is in the unit of Gauss and  $r_{adjusted}$
is in the unit of solar radii. Figure~\ref{fig:reg} shows the linear regression results.

The discrepancy between the open flux from observations and models is one
of the unsolved problems in the community \citep{Linker_2017}. \cite{Lee_2011}
have explored varying the source surface heights to match the observed open flux.
But their approach was limited to comparing the PFSS results with various 
observations. 
Our approach is different as we try to connect the magnetic
field structures between the PFSS and AWSoM. The proposed adjusted source surface
height gives larger open fluxes during the solar minimum conditions, 
which are closer to the observed open fluxes. Near
solar maximum, the PFSS model with 
adjusted source surface height gives less open flux,
which makes the comparison with the observations worse. Based on Figure\,\ref{fig:x=0}, we notice large coronal loops over strong magnetic field regions, which may bring significant open fluxes originated from active 
regions residing near the open and closed boundaries as 
suggested by \cite{Arge_2023}.

Last but not least, our work is limited to the spherical shape of the source surface for the PFSS model. Recent work has suggested that a non-spherical shape of the source surface is important during solar minimum and less important during solar maximum \citep{Kruse_2021} . It is important to investigate how the non-spherical shape evolves during a solar rotation. However, this is beyond the scope of the manuscript as our study focuses on how to adjust the height of spherical source surface based on MHD solutions. More work is needed in this direction in the future.

\begin{acknowledgments}

This work was primarily supported by the NSF PRE-EVENTS grant No. 1663800, 
the NSF SWQU grant No. PHY-2027555, 
the NSF Solar Terrestrial grant No. 2323303, 
and the NASA grants Nos. 80NSSC22K0892, 80NSSC22K0269 and 80NSSC23K0450.
J. Huang is also supported by the NASA grant 80NSSC23K0737.

We acknowledge the high-performance computing support from Cheyenne
(doi:10.5065/D6RX99HX) provided by NCAR's Computational and Information Systems Laboratory,
sponsored by the NSF, and the computation time on Frontera (doi:10.1145/3311790.3396656)
sponsored by NSF and the NASA supercomputing system Pleiades. 

This work utilizes data produced collaboratively between AFRL/ADAPT and NSO/NISP.

\end{acknowledgments}

\bibliography{reference}

\begin{thebibliography}{52}
\expandafter\ifx\csname natexlab\endcsname\relax\def\natexlab#1{#1}\fi

\bibitem[{{Alissandrakis}(1981)}]{Alissandrakis_1981}
{Alissandrakis}, C.~E. 1981, \aap, 100, 197

\bibitem[{{Altschuler} \& {Newkirk}(1969)}]{Altschuler_1969}
{Altschuler}, M.~D., \& {Newkirk}, G. 1969, \solphys, 9, 131

\bibitem[{{Antiochos} {et~al.}(2011){Antiochos}, {Miki{\'c}}, {Titov},
  {Lionello}, \& {Linker}}]{Antiochos_2011}
{Antiochos}, S.~K., {Miki{\'c}}, Z., {Titov}, V.~S., {Lionello}, R., \&
  {Linker}, J.~A. 2011, \apj, 731, 112

\bibitem[{{Arge} {et~al.}(2023){Arge}, {Leisner}, {Wallace}, \&
  {Henney}}]{Arge_2023}
{Arge}, C.~N., {Leisner}, A., {Wallace}, S., \& {Henney}, C.~J. 2023, arXiv
  e-prints, arXiv:2304.07649

\bibitem[{{Arge} \& {Pizzo}(2000)}]{Arge_2000}
{Arge}, C.~N., \& {Pizzo}, V.~J. 2000, \jgr, 105, 10465

\bibitem[{{Aschwanden} \& {Malanushenko}(2013)}]{Aschwanden_2013}
{Aschwanden}, M.~J., \& {Malanushenko}, A. 2013, \solphys, 287, 345

\bibitem[{{Badman} {et~al.}(2020){Badman}, {Bale}, {Mart{\'\i}nez Oliveros},
  {Panasenco}, {Velli}, {Stansby}, {Buitrago-Casas}, {R{\'e}ville}, {Bonnell},
  {Case}, {Dudok de Wit}, {Goetz}, {Harvey}, {Kasper}, {Korreck}, {Larson},
  {Livi}, {MacDowall}, {Malaspina}, {Pulupa}, {Stevens}, \&
  {Whittlesey}}]{Badman_2020}
{Badman}, S.~T., {Bale}, S.~D., {Mart{\'\i}nez Oliveros}, J.~C., {et~al.} 2020,
  \apjs, 246, 23

\bibitem[{{Bogdan} \& {Low}(1986)}]{Bogdan_1986}
{Bogdan}, T.~J., \& {Low}, B.~C. 1986, \apj, 306, 271

\bibitem[{{Feng} {et~al.}(2011){Feng}, {Zhang}, {Xiang}, {Yang}, {Jiang}, \&
  {Wu}}]{Feng_2011}
{Feng}, X., {Zhang}, S., {Xiang}, C., {et~al.} 2011, \apj, 734, 50

\bibitem[{{Gary}(1989)}]{Gary_1989}
{Gary}, G.~A. 1989, \apjs, 69, 323

\bibitem[{{Gombosi} {et~al.}(2021){Gombosi}, {Chen}, {Glocer}, {Huang}, {Jia},
  {Liemohn}, {Manchester}, {Pulkkinen}, {Sachdeva}, {Al Shidi}, {Sokolov},
  {Szente}, {Tenishev}, {Toth}, {van der Holst}, {Welling}, {Zhao}, \&
  {Zou}}]{Gombosi_2021}
{Gombosi}, T.~I., {Chen}, Y., {Glocer}, A., {et~al.} 2021, Journal of Space
  Weather and Space Climate, 11, 42

\bibitem[{{Gosling} \& {Pizzo}(1999)}]{Gosling_1999}
{Gosling}, J.~T., \& {Pizzo}, V.~J. 1999, \ssr, 89, 21

\bibitem[{{Groth} {et~al.}(2000){Groth}, {De Zeeuw}, {Gombosi}, \&
  {Powell}}]{Groth_2000}
{Groth}, C.~P.~T., {De Zeeuw}, D.~L., {Gombosi}, T.~I., \& {Powell}, K.~G.
  2000, \jgr, 105, 25053

\bibitem[{{Henadhira Arachchige} {et~al.}(2022){Henadhira Arachchige}, {Cohen},
  {Munoz-Jaramillo}, \& {Yeates}}]{Henadhira_2022}
{Henadhira Arachchige}, K., {Cohen}, O., {Munoz-Jaramillo}, A., \& {Yeates},
  A.~R. 2022, \apj, 938, 39

\bibitem[{{Hu} \& {Dasgupta}(2008)}]{Hu_2008a}
{Hu}, Q., \& {Dasgupta}, B. 2008, \solphys, 247, 87

\bibitem[{{Hu} {et~al.}(2008){Hu}, {Dasgupta}, {Choudhary}, \&
  {B{\"u}chner}}]{Hu_2008b}
{Hu}, Q., {Dasgupta}, B., {Choudhary}, D.~P., \& {B{\"u}chner}, J. 2008, \apj,
  679, 848

\bibitem[{{Huang} {et~al.}(2023){Huang}, {T{\'o}th}, {Sachdeva}, {Zhao}, {van
  der Holst}, {Sokolov}, {Manchester}, \& {Gombosi}}]{Huang_2023}
{Huang}, Z., {T{\'o}th}, G., {Sachdeva}, N., {et~al.} 2023, \apjl, 946, L47

\bibitem[{{Issan} {et~al.}(2023){Issan}, {Riley}, {Camporeale}, \&
  {Kramer}}]{Issan_2023}
{Issan}, O., {Riley}, P., {Camporeale}, E., \& {Kramer}, B. 2023, Space
  Weather, 21, e2023SW003555

\bibitem[{{Jian} {et~al.}(2016){Jian}, {MacNeice}, {Mays}, {Taktakishvili},
  {Odstrcil}, {Jackson}, {Yu}, {Riley}, \& {Sokolov}}]{Jian_2016}
{Jian}, L.~K., {MacNeice}, P.~J., {Mays}, M.~L., {et~al.} 2016, Space Weather,
  14, 592

\bibitem[{{Jin} {et~al.}(2012){Jin}, {Manchester}, {van der Holst},
  {Gruesbeck}, {Frazin}, {Landi}, {Vasquez}, {Lamy}, {Llebaria}, {Fedorov},
  {Toth}, \& {Gombosi}}]{Jin_2012}
{Jin}, M., {Manchester}, W.~B., {van der Holst}, B., {et~al.} 2012, \apj, 745,
  6

\bibitem[{{Kruse} {et~al.}(2021){Kruse}, {Heidrich-Meisner}, \&
  {Wimmer-Schweingruber}}]{Kruse_2021}
{Kruse}, M., {Heidrich-Meisner}, V., \& {Wimmer-Schweingruber}, R.~F. 2021,
  \aap, 645, A83

\bibitem[{{Kruse} {et~al.}(2020){Kruse}, {Heidrich-Meisner},
  {Wimmer-Schweingruber}, \& {Hauptmann}}]{Kruse_2020}
{Kruse}, M., {Heidrich-Meisner}, V., {Wimmer-Schweingruber}, R.~F., \&
  {Hauptmann}, M. 2020, \aap, 638, A109

\bibitem[{{Lee} {et~al.}(2011){Lee}, {Luhmann}, {Hoeksema}, {Sun}, {Arge}, \&
  {de Pater}}]{Lee_2011}
{Lee}, C.~O., {Luhmann}, J.~G., {Hoeksema}, J.~T., {et~al.} 2011, \solphys,
  269, 367

\bibitem[{{Levine} {et~al.}(1982){Levine}, {Schulz}, \&
  {Frazier}}]{Levine_1982}
{Levine}, R.~H., {Schulz}, M., \& {Frazier}, E.~N. 1982, \solphys, 77, 363

\bibitem[{Linker {et~al.}(2017)Linker, Caplan, Downs, Riley, Mikic, Lionello,
  Henney, Arge, Liu, Derosa, {et~al.}}]{Linker_2017}
Linker, J., Caplan, R., Downs, C., {et~al.} 2017, The Astrophysical Journal,
  848, 70

\bibitem[{{Lloveras} {et~al.}(2020){Lloveras}, {V{\'a}squez}, {Nuevo}, {Mac
  Cormack}, {Sachdeva}, {Manchester}, {Van der Holst}, \&
  {Frazin}}]{Lloveras_2020}
{Lloveras}, D.~G., {V{\'a}squez}, A.~M., {Nuevo}, F.~A., {et~al.} 2020,
  \solphys, 295, 76

\bibitem[{{McComas} {et~al.}(2002){McComas}, {Elliott}, {Gosling},
  {Reisenfeld}, {Skoug}, {Goldstein}, {Neugebauer}, \& {Balogh}}]{McComas_2002}
{McComas}, D.~J., {Elliott}, H.~A., {Gosling}, J.~T., {et~al.} 2002, \grl, 29,
  1290

\bibitem[{{Meadors} {et~al.}(2020){Meadors}, {Jones}, {Hickmann}, {Arge},
  {Godinez-Vasquez}, \& {Henney}}]{Meadors_2020}
{Meadors}, G.~D., {Jones}, S.~I., {Hickmann}, K.~S., {et~al.} 2020, Space
  Weather, 18, e02464

\bibitem[{{Miki{\'c}} {et~al.}(1999){Miki{\'c}}, {Linker}, {Schnack},
  {Lionello}, \& {Tarditi}}]{Mikic_1999}
{Miki{\'c}}, Z., {Linker}, J.~A., {Schnack}, D.~D., {Lionello}, R., \&
  {Tarditi}, A. 1999, Physics of Plasmas, 6, 2217

\bibitem[{{Neukirch}(1995)}]{Neukirch_1995}
{Neukirch}, T. 1995, \aap, 301, 628

\bibitem[{Nikoli{\'c}(2019)}]{Nikolic_2019}
Nikoli{\'c}, L. 2019, Space Weather, 17, 1293

\bibitem[{{Oran} {et~al.}(2013){Oran}, {van der Holst}, {Landi}, {Jin},
  {Sokolov}, \& {Gombosi}}]{Oran_2013}
{Oran}, R., {van der Holst}, B., {Landi}, E., {et~al.} 2013, \apj, 778, 176

\bibitem[{{Powell} {et~al.}(1999){Powell}, {Roe}, {Linde}, {Gombosi}, \& {de
  Zeeuw}}]{Powell_1999}
{Powell}, K.~G., {Roe}, P.~L., {Linde}, T.~J., {Gombosi}, T.~I., \& {de Zeeuw},
  D.~L. 1999, Journal of Computational Physics, 154, 284

\bibitem[{{Sachdeva} {et~al.}(2019){Sachdeva}, {van der Holst}, {Manchester},
  {T{\'o}th}, {Chen}, {Lloveras}, {V{\'a}squez}, {Lamy}, {Wojak}, {Jackson},
  {Yu}, \& {Henney}}]{Sachdeva_2019}
{Sachdeva}, N., {van der Holst}, B., {Manchester}, W.~B., {et~al.} 2019, \apj,
  887, 83

\bibitem[{Sachdeva {et~al.}(2021)Sachdeva, TÃ³th, Manchester, Van Der~Holst,
  Huang, Sokolov, Zhao, Al-Shidi, Chen, Gombosi, \& et~al.}]{Sachdeva_2021}
Sachdeva, N., TÃ³th, G., Manchester, W.~B., {et~al.} 2021, Earth and Space
  Science Open Archive, 13

\bibitem[{{Schatten} {et~al.}(1969){Schatten}, {Wilcox}, \&
  {Ness}}]{Schatten_1969}
{Schatten}, K.~H., {Wilcox}, J.~M., \& {Ness}, N.~F. 1969, \solphys, 6, 442

\bibitem[{{Schrijver} {et~al.}(2006){Schrijver}, {De Rosa}, {Metcalf}, {Liu},
  {McTiernan}, {R{\'e}gnier}, {Valori}, {Wheatland}, \&
  {Wiegelmann}}]{Schrijver_2006}
{Schrijver}, C.~J., {De Rosa}, M.~L., {Metcalf}, T.~R., {et~al.} 2006,
  \solphys, 235, 161

\bibitem[{{Schulz}(1997)}]{Schulz_1997}
{Schulz}, M. 1997, Annales Geophysicae, 15, 1379

\bibitem[{{Shi} {et~al.}(2024){Shi}, {Manchester}, {Landi}, {van der Holst},
  {Szente}, {Chen}, {T{\'o}th}, {Bertello}, \& {Pevtsov}}]{Shi_2024}
{Shi}, T., {Manchester}, W., {Landi}, E., {et~al.} 2024, \apj, 961, 60

\bibitem[{{Sokolov} {et~al.}(2013){Sokolov}, {van der Holst}, {Oran}, {Downs},
  {Roussev}, {Jin}, {Manchester}, {Evans}, \& {Gombosi}}]{Sokolov_2013}
{Sokolov}, I.~V., {van der Holst}, B., {Oran}, R., {et~al.} 2013, \apj, 764, 23

\bibitem[{{Szente} {et~al.}(2022){Szente}, {Landi}, \& {van der
  Holst}}]{Szente_2022}
{Szente}, J., {Landi}, E., \& {van der Holst}, B. 2022, \apj, 926, 35

\bibitem[{{Szente} {et~al.}(2023){Szente}, {Landi}, \& {van der
  Holst}}]{Szente_2023}
---. 2023, \apjs, 269, 37

\bibitem[{{Titov} {et~al.}(2011){Titov}, {Miki{\'c}}, {Linker}, {Lionello}, \&
  {Antiochos}}]{Titov_2011}
{Titov}, V.~S., {Miki{\'c}}, Z., {Linker}, J.~A., {Lionello}, R., \&
  {Antiochos}, S.~K. 2011, \apj, 731, 111

\bibitem[{{T{\'o}th} {et~al.}(2011){T{\'o}th}, {van der Holst}, \&
  {Huang}}]{Toth_2011}
{T{\'o}th}, G., {van der Holst}, B., \& {Huang}, Z. 2011, \apj, 732, 102

\bibitem[{{T{\'o}th} {et~al.}(2005){T{\'o}th}, {Sokolov}, {Gombosi}, {Chesney},
  {Clauer}, {de Zeeuw}, {Hansen}, {Kane}, {Manchester}, {Oehmke}, {Powell},
  {Ridley}, {Roussev}, {Stout}, {Volberg}, {Wolf}, {Sazykin}, {Chan}, {Yu}, \&
  {K{\'o}ta}}]{Toth_2005}
{T{\'o}th}, G., {Sokolov}, I.~V., {Gombosi}, T.~I., {et~al.} 2005, Journal of
  Geophysical Research (Space Physics), 110, 12226

\bibitem[{{T{\'o}th} {et~al.}(2012){T{\'o}th}, {van der Holst}, {Sokolov}, {De
  Zeeuw}, {Gombosi}, {Fang}, {Manchester}, {Meng}, {Najib}, {Powell}, {Stout},
  {Glocer}, {Ma}, \& {Opher}}]{Toth_2012}
{T{\'o}th}, G., {van der Holst}, B., {Sokolov}, I.~V., {et~al.} 2012, Journal
  of Computational Physics, 231, 870

\bibitem[{{van der Holst} {et~al.}(2019){van der Holst}, {Manchester}, {Klein},
  \& {Kasper}}]{vanderholst_2019}
{van der Holst}, B., {Manchester}, W.~B., I., {Klein}, K.~G., \& {Kasper},
  J.~C. 2019, \apjl, 872, L18

\bibitem[{{van der Holst} {et~al.}(2014){van der Holst}, {Sokolov}, {Meng},
  {Jin}, {Manchester}, {T{\'o}th}, \& {Gombosi}}]{vanderholst_2014}
{van der Holst}, B., {Sokolov}, I.~V., {Meng}, X., {et~al.} 2014, \apj, 782, 81

\bibitem[{{van der Holst} {et~al.}(2022){van der Holst}, {Huang}, {Sachdeva},
  {Kasper}, {Manchester}, {Borovikov}, {Chandran}, {Case}, {Korreck}, {Larson},
  {Livi}, {Stevens}, {Whittlesey}, {Bale}, {Pulupa}, {Malaspina}, {Bonnell},
  {Harvey}, {Goetz}, \& {MacDowall}}]{vanderholst_2022}
{van der Holst}, B., {Huang}, J., {Sachdeva}, N., {et~al.} 2022, \apj, 925, 146

\bibitem[{Wang {et~al.}(2000)Wang, Lean, \& Sheeley~Jr}]{Wang_2000}
Wang, Y.-M., Lean, J., \& Sheeley~Jr, N. 2000, Geophysical Research Letters,
  27, 505

\bibitem[{{Wang} \& {Sheeley}(1990)}]{Wang_1990}
{Wang}, Y.~M., \& {Sheeley}, N.~R., J. 1990, \apj, 355, 726

\bibitem[{{Wang} \& {Sheeley}(1992)}]{Wang_1992}
---. 1992, \apj, 392, 310

\end{thebibliography}

\end{document}